\documentstyle[11pt,fleqn,epsf]{article}
\newenvironment{eq}{
	\setlength{\mathindent}{0 cm}\begin{equation}}{\end{equation}}
\newcommand{\Integer}{\:\mbox{\sf Z} \hspace{-0.82em} \mbox{\sf Z}\,}
\def\Mult#1#2#3{\left[#1 \atop #2 \right]_{#3}}
\def\Mults#1#2#3{\left[{\textstyle {#1 \atop #2} } \right]_{#3}}

\begin{document}

\title{
A bijection which implies Melzer's polynomial identities: \\
the $\chi_{1,1}^{(p,p+1)}$ case}

\author{
Omar Foda\thanks{
e-mail: {\tt foda@mundoe.maths.mu.oz.au}} \ and S.~Ole Warnaar\thanks{
e-mail: {\tt warnaar@mundoe.maths.mu.oz.au}} \\
{\it Department of Mathematics,}  \\
{\it The University of Melbourne,} \\
{\it Parkville, Victoria 3052,}  \\
{\it Australia.}}

\date{January, 1995 \\ \hspace{1mm}
\\
Preprint No. 03-95}

\maketitle

\begin{abstract}
We obtain a bijection between certain lattice paths and
partitions. This implies a proof of polynomial
identities conjectured by Melzer. In a limit, these
identities reduce to Rogers--Ramanujan-type identities for
the $\chi_{1,1}^{(p,p+1)}(q)$ Virasoro characters, conjectured
by the Stony Brook group.
\end{abstract}

\section{Introduction}
In \cite{Melzer}, Melzer conjectured polynomial
identities that include the following:
\begin{eqnarray}
\lefteqn{
\sum_{m_1,\ldots,m_{p-2}= 0}^{\infty}
q^{\displaystyle  \vec{m}^T C_{p-2} \: \vec{m} }
\prod_{j=1}^{p-2}
\Mult{ m_{j-1} + m_{j+1}}{2 m_j}{q} }
\nonumber \\ & & \nonumber \\
&& =
\sum_{j=-\infty}^{\infty} \left\{
q^{p(p+1)j^2+j}
\Mult{2m}{m-(p+1)j}{q}
-q^{p(p+1)j^2+(2p+1)j+1}
\Mult{2m}{m-1-(p+1)j}{q} \right\}.
\label{PolyId}
\end{eqnarray}
Here $m_0\equiv m$, $m_{p-1}\equiv 0$,
$\vec{m}^T = (m_1,\ldots,m_{p-2})$,
$C_{p-2}$ is the Cartan matrix of the Lie algebra A$_{p-2}$:
\begin{equation}
(C_{p-2})_{j,k} = 2\delta_{j,k}-\delta_{j,k+1}-\delta_{j,k+1},
\qquad j,k=1,\ldots,p-2,
\end{equation}
and $\Mults{N}{m}{q}$ is the $q$-binomial coefficient:
\begin{equation}
\Mult{N}{m}{q} =  \left\{
\begin{array}{ll}
\displaystyle \frac{(q)_N}{(q)_m (q)_{N-m}} \quad & 0\leq m \leq N \\
& \\
0 & otherwise,
\end{array} \right.
\end{equation}
with $(q)_m = \prod_{k=1}^m (1-q^k)$ for $m>0$ and $(q)_0=1$.
In the limit $m\to \infty$, the above identities reduce to
Rogers--Ramanujan--type identities for the $\chi_{1,1}^{(p,p+1)}(q)$
Virasoro characters, conjectured by the Stony Brook group
in \cite{KKMM}. For an introduction to these identities,
and further references, we refer the reader to \cite{DKKMM}.

In \cite{Berkovich,Warnaar} proofs of the above identities obtained.
In this letter, we present a bijective proof based on the following
observation: The LHS of (\ref{PolyId}) is the generating function
of certain {\em restricted lattice paths} of finite lengths
\cite{Warnaar}. On the other hand, the RHS is the generating
function of partitions with prescribed hook differences
\cite{ABBBFV}. In this work, we establish a bijection
between these lattice paths and partitions.

The bijection that we obtain is {\em so} simple, it almost
trivializes the identities (given that we know how to evaluate
the generating functions of the combinatoric objects that appear
on each side: the lattice paths, and the Ferrers graphs, and this
is {\em not} a trivial task). We hope that this work will stimulate
the search for further bijections that equally simply imply the
rest of Melzer's and other polynomial identities.
Since bijections contain {\em detailed} information about the
objects involved, we hope that this and similar proofs help
gain better understanding of the full meaning of character
identities.

\vspace{5 mm}

In \S 2, we define the lattice paths generated by the RHS of
(\ref{PolyId}), and the partitions generated by the LHS.
In \S 3, we establish a bijection between these lattice paths
and partitions, using certain matrices as interpolating
structures. \S 4 contains a short discussion.

\section{Paths and partitions}

\subsection{Admissible lattice paths}

We define an {\em admissible sequence of integers
$\Sigma$} as an ordered sequence
$\{\sigma_0,\sigma_1,\sigma_2,\ldots,\sigma_{2m}\}$,
with $|\sigma_j-\sigma_{j+1}|=1$, for all $j$,
$\sigma_0=\sigma_{2m}=0$,
and each $\mbox{$\sigma_j\in \{0,1,\ldots,p-1\}$}$.

Given $\Sigma$, we obtain {\em an admissible lattice
path} $P(\Sigma)$ as follows: plot $(j,\sigma_j)$,
$j=0,\ldots,2m$ in $R^2$, and connect each pair of adjacent
vertices $(j,\sigma_j)$ and $\mbox{$(j+1,\sigma_{j+1})$}$ by
a straight line-segment.
An example of $P$ with $p\geq 6$ is shown in Fig.~1.
For the rest of this work, the term {\em path} will always
be used to imply an admissible lattice path
in the above sense.

\subsubsection{Vertices on a path}

On a path, we distinguish four different types of
vertices:

\begin{description}
\item[V1] If $\sigma_{j-1}<\sigma_j<\sigma_{j+1}$, then $(j,\sigma_j)$
   is an up-vertex,
\item[V2] if $\sigma_{j-1}>\sigma_j>\sigma_{j+1}$, then $(j,\sigma_j)$
   is a down-vertex,
\item[V3] if $\sigma_{j-1}<\sigma_j>\sigma_{j+1}$, then $(j,\sigma_j)$
   is a maximum,
\item[V4] if $\sigma_{j-1}>\sigma_j<\sigma_{j+1}$, then $(j,\sigma_j)$
   is a minimum.
\end{description}
By definition, the end-points $(0,0)$ and $(2m,0)$ are both
minima. Notice that a path consists of a number of sections,
each connecting a maximum and an adjacent minimum. Since each
path begins and ends on the $x$-axis, there is always an even
number of such sections, with an equal number of {\em ascending}
and {\em descending} sections, as one scans the paths from one
end to the other. The total number of these sections is twice
the number of maxima. Furthermore, there is always an equal number
of up- and down-vertices, and one more minimum than maximum. Example:
in Fig.~1, we find 14 sections, 6 up- and down-vertices, 7 maxima
and 8 minima adding up to $27=2m+1$.

\subsubsection{The weight $W(P)$ of a path}

To a path $P$ we assign the {\em weight} $W(P)$ as half the sum of
the $x$-coordinates of its up- and down-vertices. For the example
of Fig.~1, we obtain the weight $(1+2+4+6+7+8+10+11+14+15+17+21)/2=58$.
The generating function $F$ of our paths is defined as
\begin{equation}
F(q) = \sum_{P} q^{W(P)}.
\end{equation}

\begin{description}
\item{\bf Theorem \cite{Warnaar}:}
For all $p \in \Integer_{\geq 2}$ and $m\in \Integer_{\geq 0}$ we have
\begin{equation}
F(q) =
\sum_{m_1,\ldots,m_{p-2}= 0}^{\infty}
q^{\displaystyle  \vec{m}^T C_{p-2} \: \vec{m} }
\prod_{j=1}^{p-2}
\Mult{m_{j-1}+m_{j+1}}{2 m_j}{q} ,
\end{equation}
\end{description}

For proof, we refer the reader to \cite{Warnaar}.
Note that the case $p=2$ is trivial, as we have only
a single path without any up- or down-vertices,
and $F=1$.

\subsection{Partitions with prescribed hook differences}

For completeness, we list a number of standard definitions
from the theory of partitions \cite{Andrews,ABBBFV}.

A {\em partition} of $n\in \Integer_{>0}$ is a sequence of
integers $\{r_1,r_2,\ldots,r_N\}$, with $\sum_{j} r_j = n$ and
$r_1\geq r_2 \geq \ldots \geq r_N >0$ \cite{Andrews}. $r_1$
and $N$ are the largest part and the number of parts of the
partition, respectively. Example: a partition of $58$ is
$11,9,9,8,8,7,4,1,1$. Here, the largest part is 11, and the
number of parts is 8.

The {\em Ferrers graph} corresponding to a partition is
obtained by drawing $N$ rows of nodes, with the $j$-th
row containing $r_j$ nodes, counting rows from top to bottom.
The Ferrers graph of the above partition of 58 is shown in
Fig.~2a. The largest square of nodes one can draw in a
Ferrers graph is called the {\em Durfee square}. The dimension
$D$ of the side of a Durfee square is the number of $r_j\geq j$.

The {\em length} of the $j$-th row/column of a Ferrers graph is
the number of nodes of the $j$-th column/row, counting
columns from left to right. The length of the $j$-th row being
$r_j$, the length of the $j$-th column will be denoted $c_j$.

The $(j,k)$-th node of a Ferrers graph is the node in the
$j$-th row and $k$-th column. The nodes with coordinates
$(k+j,k)$ form the {\em $j$-th diagonal}. A quantity that
will play an important role is this work is the {\em hook
difference at node} $(j,k)$, defined as $r_j-c_k$

The Ferrers graph of the above partition of 58 is shown in
Fig.~2a. In this same figure, we also show the Durfee square
and indicate some of the diagonals.

\subsubsection{Admissible partitions}

The partitions relevant to (\ref{PolyId}) satisfy two
conditions. In the language of Ferrers graphs, these
are:
\begin{description}
\item[F1] The \lq boundary conditions\rq: the number of columns
$ \leq m$, and the number of rows is $\leq m - 1$.
\item[F2] The hook differences: on the zeroth diagonal they
      satisfy  $\geq 1$, on the $(p-2)$-th diagonal they
      satisfy $\leq 0$.
\end{description}

For the rest of this work, we use {\em partition} to
mean an admissible partition that satisfies the above
conditions. For the partition of Fig.~2a we have computed
the hook differences on the $0,3$ and $4$-th diagonal in
Fig.~2b. We hence find admissibility for $p\geq 6, m\geq 11$.

The generating function $B$ of our partitions is defined as
\begin{equation}
B(q) = \sum_{n=0}^{\infty} G\left(\Pi(n)\right) q^n,
\end{equation}
with $G(n)$ the number of partitions of $n$.

\begin{description}
\item{\bf Theorem \cite{ABBBFV}:}
For all $p \in \Integer_{\geq 2}$ and $m\in \Integer_{\geq 0}$ we have
\begin{eq}
B(q) =
\sum_{j=-\infty}^{\infty} \left\{
q^{p(p+1)j^2+j}
\Mult{2m}{m-(p+1)j}{q}
-q^{p(p+1)j^2+(2p+1)j+1}
\Mult{2m}{m-1-(p+1)j}{q} \right\},
\label{B}
\end{eq}
\end{description}
For proof, we refer the reader to \cite{ABBBFV}.

As for the generating function $F$ of the lattice paths,
we note that for $p=2$ things trivialize. Clearly, from
our definition of admissible partitions, for $p=2$ we
obtain $B=1$, as the restrictions on the $0$-th diagonal
are exclusive. (The generating function of the partition
of $0$ is 1, by definition.) To observe that expression
(\ref{B}) indeed yields unity one has to invoke the finite
analogue of Euler's identity \cite{Andrews}.

\section{A bijection between paths and partitions}

\subsection{Interpolating matrices}

Since a path is uniquely defined in terms of the ordered
sequence of up- and down-vertices that it contains, we
can encode it in terms of a matrix whose elements are
precisely such a sequence. Such a matrix will interpolate
between a path and the corresponding partition.

\subsection{From paths to matrices}

A path $P$ can be encoded into a 2--by--$\cal N$ matrix
$M(P)=\left(
\begin{array}{cccc}
u_1 & u_2 & \ldots & u_{\cal N} \\
d_1 & d_2 & \ldots & d_{\cal N}
\end{array} \right)$,
with $\cal N$ the number of maxima of $P$, as follows:
$u_j$ is the number of up-vertices between the $j$-th minimum
and the $j$-th maximum, and $d_j$ is the number of down-vertices
between the $j$-th maximum and the $(j+1)$-th minimum. Here we
count minima and maxima from left to right, $(0,0)$ being the
first minimum.  Example: for the path of Fig.~1, we obtain
$M=\left(
\begin{array}{ccccccc}
2 & 3 & 0 & 1 & 0 & 0 & 0 \\
1 & 2 & 2 & 0 & 1 & 0 & 0
\end{array}
\right)$.

\subsubsection{Restrictions on the interpolating matrices}

The restrictions on the paths translate into restrictions
on the interpolating matrices as follows:

\begin{enumerate}
\item[R1] Since a path begins and ends on the real axis,
the elements of the corresponding matrix satisfy

$$
\sum_{j=1}^{\cal N} u_j = \sum_{j=1}^{\cal N} d_j
$$

\item[R2] Since the paths have length $2m$, the total
number of {\em bonds} on a path is $2m$. In any section
between a maximum and a minimum, the number of
bonds exceeds the number of vertices by one.
Since the elements of the
interpolating matrices are precisely the number of vertices
on such sections, they satisfy

$$
2{\cal N} + \sum_{j=1}^{\cal N} (u_j + d_j)= 2m
$$

\item[R3] The condition that paths are confined to the strip
$0\leq y \leq p-1$ gives
\begin{equation}
\begin{array}{l}
\displaystyle \sum_{k=1}^j (u_k-d_k) \geq 0, \\
\displaystyle \sum_{k=1}^j (u_k-d_{k-1}) \leq p-2,
\end{array} \qquad \qquad
j=1,\ldots,{\cal N}, \qquad  d_0\equiv 0.
\label{iii}
\end{equation}

\end{enumerate}

We refer to any 2--by--${\cal N}$ matrix of non-negative integer
elements which satisfy conditions R1--R3 above, as {\em admissible}.
The point is that one can always use the elements of such a matrix
to uniquely construct {\em a} path. However, the restrictions on the
elements of an admissible matrix translate to the restrictions satisfied
by a path. For the rest of this work, the term {\em matrix} will be
used to imply an admissible interpolating matrix.

\subsubsection{From matrices to paths}

Since, as explained above, an admissible interpolating matrix
simply encodes the ordered sequences of up- and down-vertices,
it can be used to uniquely construct an admissible  path, and
the map from paths to matrices, as above, is a bijection.

\subsubsection{The weight of a path}

Given the matrix $M(P)$ corresponding to path $P$, the
weight, $W(P)$, of the latter--which was introduced
in \S 2.1--can be computed as
\begin{eqnarray}
W(P) &=& \frac{1}{2} \left(
\sum_{k=1}^{u_1} k +
\sum_{k=1}^{d_1} (1+u_1+k) +
\sum_{k=1}^{u_2} (2+u_1+d_1+k)  \right. \nonumber \\
& &  \qquad \qquad \qquad \qquad \qquad   \left. +
\sum_{k=1}^{d_2} (3+u_1+d_1+u_2+k) + \ldots \right).
\label{wp}
\end{eqnarray}
Using restriction R1 on the elements of $M(P)$, and
defining the number $D$ by
\begin{equation}
D \equiv \sum_{j=1}^{\cal N} u_j = \sum_{j=1}^{\cal N} d_j,
\label{defD}
\end{equation}
we can rewrite (\ref{wp}) as
\begin{equation}
W(P) = D(D+1) + \sum_{j=1}^{\cal N} \left[
\sum_{k=j+1}^{\cal N} u_k + \sum_{k=j+1}^{\cal N} d_k \right].
\label{wp2}
\end{equation}
This form of the weight function will be useful later. Basically,
we will translate it directly into a Ferrers graph of a partition
of $W(P)$.

\subsection{A bijection between matrices and partitions}

Given an admissible matrix $M(P)$, with a weight $W(P)$ given
by equation (\ref{wp2}), we construct a Ferrers graph of the number
$W(P)$ as follows:

\subsubsection{From matrices to partitions}

\begin{description}
\item[P1]  We translate the term $D(D+1)$ to a rectangle of nodes,
which has $D$ rows and $(D+1)$ columns. This is basically
a Durfee square, plus an extra column of length $D$, of
some Ferrers graph. Let us refer to this as a {\em Durfee
rectangle}. In the following, we generate the rest
of the graph from the other two terms in (\ref{wp}).

\item[P2] The lengths of the $(D+j)$-th row (below the
Durfee rectangle constructed above) and the $(D+j+1)$-th
column (to the right of the Durfee rectangle) are given by
\begin{equation}
\begin{array}{rll}
r_{D+j}  &= \displaystyle\sum_{k=j+1}^{\cal N} u_k, & \qquad
j=1,\ldots, {\cal N}, \\
c_{D+j+1} & = \displaystyle\sum_{k=j+1}^{\cal N} d_k, & \qquad
j=1,\ldots, {\cal N}.
\end{array}
\label{rc}
\end{equation}
\end{description}
We note that P1 and P2 only make sense provided $D$ is
defined
as in (\ref{defD}). Hence in our construction of partitions
we have implicitly used condition R1 on admissible matrices.
This also
implies the extra equation
\begin{equation}
u_1 = D - r_{D+1}.
\end{equation}
needed to invert the transformation from $M(P)$ to $W(P)$.
We also note that, from (10), $j=0$ in the first equation in (\ref{rc})
gives the dimension of the side of the Durfee square: $D$, which
is already constructed in step P1 above. similarly, $j=0$ in the
second equation in (\ref{rc}) gives the right-most column of the
Durfee rectangle, which is also already constructed.


The above construction uniquely generates a Ferrers graph, which
has a Durfee rectangle of dimensions $D$--by--$(D+1)$. Since the
total number of nodes of a Ferrers graph is
\begin{equation}
D^2 + \sum_{j> D} (r_j + c_j),
\end{equation}
it follows from (\ref{wp2}) that (\ref{rc}) indeed defines
a partition of the number $W(P)$.

Next, we have to show that the Ferrers graph generated, as
explained above, are indeed what we are looking for. In
other words, it has to satisfy the conditions F1 and F2 of \S 2.2.

\subsubsection{The partitions obtained are admissible}

First, we consider the conditions R1 and R2. Setting
$j=\cal N$ in (\ref{rc}) we obtain
$r_{D+\cal N}=c_{D+{\cal N}+1}=0$. However, from R1 and R2 we
get $N+D=m$,  yielding
$r_m=c_{m+1}=0$. Thus, via (\ref{rc}),
each admissible matrix defines a Ferrers graph
which has maximally
$(m-1)$ rows and $m$ columns. Concluding, restrictions R1 and R2
on the admissible matrices imply the restriction F1 on
admissible partitions.

Next, restriction R3, as formulated in equation (\ref{iii}),
can be translated into conditions on the rows and columns
of the Ferrers graph as follows:
\begin{eqnarray}
\lefteqn{
\sum_{k=1}^j (u_k -d_k) = \sum_{k=j+1}^{\cal N} (d_k-u_k)
= c_{D+j+1}-r_{D+j} \geq 0, }  \\
\lefteqn{
\sum_{k=1}^j (u_k -d_{k-1}) = \sum_{k=j+1}^{\cal N} (d_{k-1}-u_k)
= c_{D+j}-c_{D+\cal N}-r_{D+j} \leq p-2, }
\end{eqnarray}
for all $j=1,\ldots,\cal N$.
The length of the $j$-th row (column) $j=1,\ldots,D$
is $D+x_j$ ($D+y_j$) with $x_j$ ($y_j$) the smallest integer
such that $c_{D+x_j+1}<j$ ($r_{D+y_j+1}<j$), which implies
$c_{D+x_j}>c_{D+x_j+1}$ ($r_{D+y_j}>r_{D+y_j+1}$).
We thus find that the hook differences on the $0$-th diagonal
are $\geq 1$, and on the $(p-2)$-th diagonal are $\leq 0$.

We have therefore mapped admissible matrices on Ferrers graphs
of admissible partitions.

\subsubsection{From partitions to matrices}

The above map is invertible since the dimensions of the
building blocks of an admissible Ferrers graphs, namely
the Durfee square, and the rows and columns adjacent to
it, can be translated to elements of an admissible matrix
using the same equations as above.

Finally, we need to show that the restrictions on
the admissible Ferrers graphs translate correctly
to the restrictions on admissible matrices.

\subsubsection{The matrices obtained are admissible}

\begin{description}

\item[M1] Restriction R1 is trivially satisfied

\item[M2] Restriction R2 can always be satisfied,
      since it basically amounts to adding extra columns
      with zero entries to the matrix until $D+{\cal N}=m$
      is satisfied

\item[M3] Restriction R3 is satisfied given that
      we start from an admissible partition.

\end{description}
This completes our bijection between paths and partitions.
As a consequence of that, the corresponding generating
functions--as obtained in \cite{Warnaar,ABBBFV}--are equal,
which implies (\ref{PolyId}).

\vspace{5mm}

Before concluding we note the following simple graphical way
to construct the Ferrers graph of an admissible
partition from an admissible matrix.
\begin{itemize}
\item
Draw a partition of $2 {\cal N}$ rows,
$r_j = 2 {\cal N}+1-j$, $r_{j+N}={\cal N}-j$,
$j=1,\ldots {\cal N}$.
\item
Multiply the $j$-th row $d_{{\cal N}-j+1}$ times and
the $j$-th column $u_{{\cal N}-j+1}$ times, $j=1,\ldots,
{\cal N}$.
\end{itemize}
Carrying out both steps result in the admissible partition.
For the partition of
Fig.~2 this is shown in Fig.~3. We thus find that the
path of Fig.~1 corresponds to the partition of Fig.~2.
The reverse of this statement is true provided $m=13$.

\section{Discussion}

In this letter, we established a bijection between a class of
restricted lattice paths and a class of partitions with
prescribed hook differences. As a result of this, we obtain
a bijective proof a Melzer's polynomial identities related
to the Virasoro characters $\chi_{1,1}^{(p,p+1)}(q)$.

We restricted ourselves to lattice paths confined to
$0\leq y \leq p-1$, starting in $(0,0)$ and ending in $(2m,0)$.
We expect that a generalization of the above bijection exists,
such that one can map lattice paths starting confined to this same strip,
but starting in $(0,s-1)$ and ending in $(M,r-1)$, to partitions
with hook differences on the $(1-r)$-th diagonal $\geq r-s+1$ and
on the $(p-r-1)$-th diagonal $\leq r-s$. If so, this
would amount to a bijective method of proof for all polynomial
$\chi_{r,s}^{(p,p+1)}(q)$ character identities \cite{Melzer}.

\subsection*{Acknowledgements}
We wish to thank Professor G.~E.~Andrews for correspondence,
and his interest in this work.
This work is supported by the Australian Research Council.

\newpage

\begin{figure}[hbt]
\centerline{\epsffile{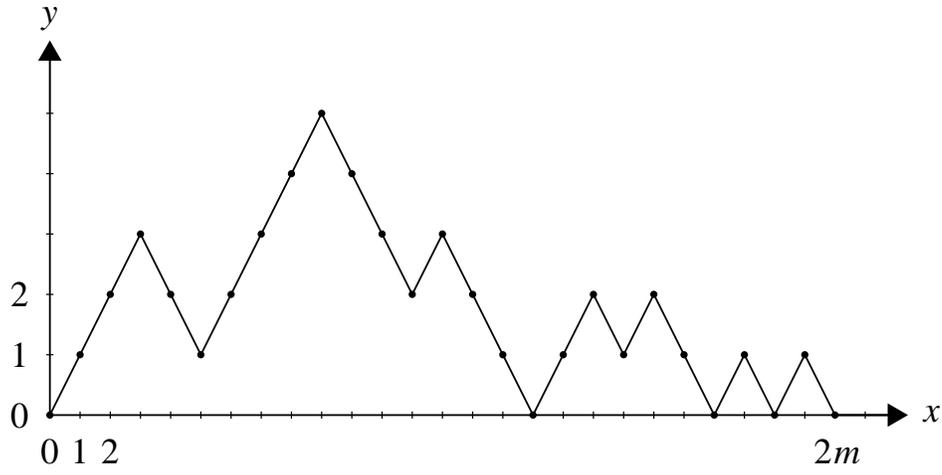}}
\caption{Example of a restricted lattice path for $p\geq 6$.}
\label{fig1}
\end{figure}

\vspace*{5mm}

\begin{figure}[hbt]
\centerline{\epsffile{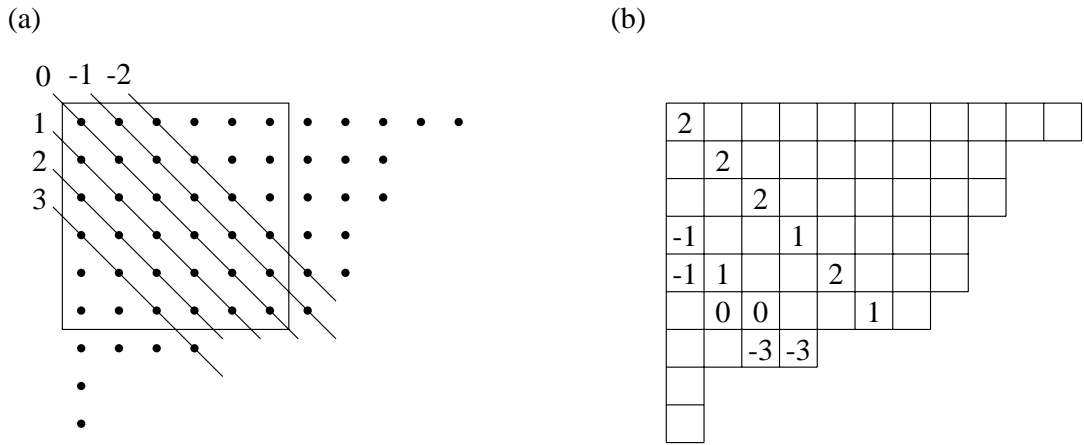}}
\caption{(a) Ferrers graph of the partition 11,9,9,8,8,7,4,1,1.
The Durfee square has size $D=6$.
(b) Young representation of the partition, with listing of hook
differences on the 0-th, 3-th and 4-th diagonal.}
\label{fig2}
\end{figure}

\vspace*{5mm}

\begin{figure}[hbt]
\centerline{\epsffile{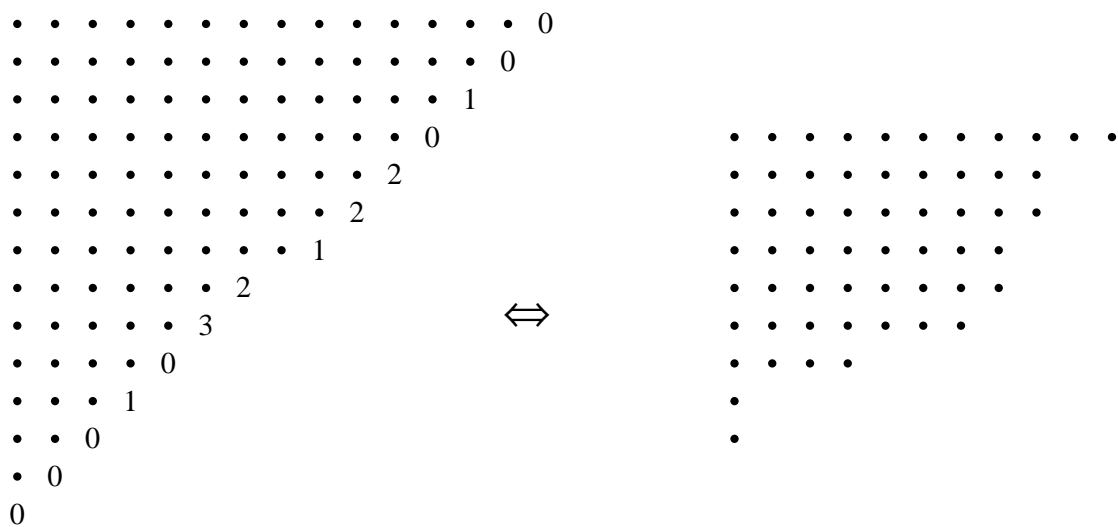}}
\caption{The construction of a Ferrers graph out of the
admissible matrix
$M=\left(
{u_1 \; u_2 \; \ldots \; u_{\cal N} \atop
 d_1 \; d_2 \; \ldots \; d_{\cal N}}
\right)=\left(
{2 \; 3 \; 0 \; 1 \; 0 \; 0 \; 0 \atop
 1 \; 2 \; 2 \; 0 \; 1 \; 0 \; 0 }
\right)$.
Clearly, from the resulting graph and the value of $m$
one can easily reconstruct the ``pre-graph''. Hereto we recall
that the dimension $D$ of the Durfee square is
$\sum u_j = \sum d_j$ and that $N=m-D$.}
\label{fig3}
\end{figure}

\end{document}